\documentclass[a4paper,12pt]{article}
\usepackage{amsmath}
\usepackage[psamsfonts]{amssymb}
\usepackage[mathscr]{eucal}
\usepackage[dvips]{graphicx}
\begin{document}

\begin{center}

\section*{  On the group-theoretical approach to energy quantization of a perturbed vortex ring: \\ spectrum calculating in the pipe-type domain.}

\vspace{3mm}

{ S.V. Talalov }

\vspace{3mm}

  {\small Dept. of Applied Mathematics,   Togliatti State University, \\ 
 14 Belorusskaya str.,  Tolyatti,     445020 Russia.     }\\
svt\_19@mail.ru

\end{center}

\begin{abstract}
In this study, the problem of  the energy spectrum of a quantum vortex loop moving in a thin long pipe is solved for the first time. 
  		We quantize this dynamic system using a new method, which leads to non-trivial results for circulation $\Gamma$ and energy values $E$.  
			It is shown that the spectrum has a quasi-continuous   fractal structure.
									In the final form, we present the spectrum of the vortex loop in the form of a ''Regge trajectory'' $E = E(\Gamma)$.  			
					 					The vortex quantization problem is considered outside of two-fluid hydrodynamics and other conventional approaches.
										We also discuss ways to improve the model, which could allow us to apply the results we've obtained to describe a quantum turbulent flow.		
										
		\end{abstract}
		

	{\bf keywords:} local induction approximation; quantum vortex filament;  extended Galilei group.   


	\section{Introduction}
	 
	\paragraph~
	
	The important role of vortices in the formation of a turbulent flow, both classical and quantum, is currently an established fact. A large number of articles have been devoted to this issue since Feynman's pioneering work \cite{Feyn}.
		These ideas stimulated numerous studies, a certain stage of which was reflected in the books \cite{Donn,Frisch}.	
				Quantum effects add complexity to the study of this problem: so, the differences between the classical and quantum turbulence were investigated in the paper \cite{Vinen}.

			According to this concept,	a single vortex is an ''atom'' of a turbulent flow. In this regard, the study of its structure and properties is of considerable interest.	
				  The crucial example here  is quantization of the circulation $\Gamma$ of a vortex filament in a quantum fluid.
										  Starting with the old  Onsager's ideas,  this procedure is performed according to the rule
				\begin{equation}
				\label{Gamma_Ons}
				\Gamma = \frac{\hbar n}{\mu_H}\,, \qquad  n = 1,2,\dots
				\end{equation}
		where $	\mu_H$ is a ''helium mass'' (for the case of superfluid helium). 	Despite decades of theory development, this formula is still used today. 
				However, the contradictions that this formula contains were discussed in the work \cite{Donn} (see section 2.3.1 of this book) many years ago.  					
		It seems that these contradictions can be a driving force behind the development of new models in this field. 	One such approach is the author's model, the development of which is the focus of this paper.  For instance, in the paper \cite{Tal_PhRF23}, arguments are presented in favor of generalizing  Eq.(\ref{Gamma_Ons}).	  						
Note that formula (\ref{Gamma_Ons}) has been experimentally tested  in the simplest cases of a linear single vortex filament only.	Numerical simulation of quantized circulation  has been investigated also 	in the paper \cite{MuPoKr} in frame of Gross - Pitaevskii model. 			
				
				One of the key moments in this area is the problem of the origin and subsequent development of turbulence. When looking at a quantum  turbulent flow as a tangle of vortices,  such events as an appearance and subsequent decay of a single vortex  arise  naturally.  
		Of course, in an equilibrium turbulent flow, it is impossible to track every such event. Statistical description of physical quantities seems to be the main method here. 
			As a result, there is a problem of correctly determining the energy of a single quantum vortex. 
								Among other things, this problem arises both because of the small size of some individual vortices in a turbulent flow and because of the small size of the vortex core. Indeed  {\it''\dots  considering how small the vortex core in helium II is, i.e., of order an angstrom, it would seem that one either ought to know how it is constructed or one ought to find a way to ignore it. Unfortunately neither goal has been achieved''} \cite{Donn}.

 In this paper we quantize certain single vortex loop and calculate spectra of circulation $\Gamma$ and energy $E$. Our studies is based on the previous author's works
\cite{Tal_EJM,Tal_PoF,Tal_PhRF23} which were devoted to the non-standard quantum description of the closed vortex filaments.
	The author's approach is motivated by the fact that quantization of a complex dynamical system is  an ambiguous procedure.
	The development of quantum theory really confirms Dirac's famous words:
	{\it''\dots methods of quantization are all of the nature of practical rules, whose application depends on consideration of simplicity''} \cite{Dirac}. For instance, the quantization of canonically equivalent but different sets of classical variables can lead to physically different results at the quantum level.
	This is true even for  simple systems from a    text-book: so,	
	the  quantization    of      a harmonic oscillator in terms of action-angle variables  demonstrates many unexpected and interesting results \cite{Kast}.
	As a rule, the problem  arises when we quantize a complex dynamic system, especially
	 in the cases where experiments cannot confirm or disprove the final results for the spectrum of observable values.
	Of course, the theory of quantum vortices has been around for decades. A lot of articles are devoted to this issue.
	Here we note only some  fundamental books of \cite{Donn,Nemir,Sonin}  that reflect different periods of research and approaches to the problem.
	
	Our study is devoted to the quantum description of the perturbed vortex ring of radius $R$.  This ring (vortex filament)   moves in the domain $D \subset {\sf R}_3$ that is defined by the conditions
	\begin{equation}
	\label{domain}
	\rho ~\le~ R_1\,, \qquad -\frac{L}{2} ~<~ z ~<~ \frac{L}{2}, \qquad\qquad R_1 << L\,, \quad R < R_1\,.
	\end{equation}
	Of course, the proposed approach will be valid for other domains as well. We chose the domain (\ref{domain}) because the spectral problem for the energy is solved explicitly in this case.  
	We use  	cylindrical coordinate system $(\rho, \varphi, z)$    with corresponding  basis $(\,\boldsymbol{n}_\rho\,, \boldsymbol{n}_\varphi\,,  \boldsymbol{n}_z\,)$ which is associated with  domain (\ref{domain})  naturally. The $z$ axis   ($ \boldsymbol{n}_z$) is directed along the axis of the cylinder under consideration. 
		In fact, we will consider the asymptotic  behavior $L\to \infty$, taking into account that the value $L < \infty$ in reality (the case of a ''long pipe''). 
		For our further studies, we use a representation for  smooth closed filament of length $S$   in the form
\begin{equation}
        \label{proj_r}
                                   {\boldsymbol r}(\tau,\xi) ~=~  {\boldsymbol{q}} + 
         {R} \int\limits_{0}^{2\pi}  \left[\, {\xi - \eta}\,\right] {\boldsymbol j}(\tau,\eta) d\eta\,, \qquad  R = \frac{S}{2\pi} \,, \qquad \xi = \frac{s}{R}\,,
                  \end{equation}
where parameter $\tau$  describes the curve evolution.  Vector ${\boldsymbol{q}} = {\boldsymbol{q}}(\tau)$ defines the position of the filament  (for example,  the   center of the circular vortex ring) in some coordinate system.
The notation $[\,x\,]$ means the integer part of the number $x/{2\pi}$:
\begin{equation}
        \label{int_part}
 [\,0\,] = 0\,, \qquad  [\,x+2\pi\,] = [\,x\,]+1\,, \qquad \forall\,x\,.
\end{equation}
 Vector ${\boldsymbol j}(\tau,\eta)$ is the unit  tangent (affine) vector for this filament. 
To ensure the periodicity of the function 
 ${\boldsymbol r}(\cdot,\xi)$ the following constraints must be fulfilled:
 \begin{equation}
        \label{con_j}
		\int\limits_{0}^{2\pi}   {\boldsymbol j}(\tau,\eta) d\eta	= 0\,.	
\end{equation}
 We use Local Induction  Approximation  (LIA) for our subsequent studies (see \cite{Saffm} for example) . This approximation leads to the equation

 \begin{equation}
\label{LIE}
\partial_\tau {\boldsymbol{r}}(\tau ,\xi)
         ~=~ \frac{\alpha}{R}\,\partial_\xi{\boldsymbol{r}}(\tau ,\xi)
				\times\partial_\xi^{\,2}{\boldsymbol{r}}(\tau ,\xi)\,. 
\end{equation}
  Let us note that  Local Induction  Approximation provides the specific description of a single vortex.  Of course, we cannot use it to describe vortices in a developed turbulent flow.
In the paper \cite{Tal_24j}, the author suggested modifying equation (\ref{LIE}).  The proposed modification is based on using the integrals of motion of the AKNS hierarchy to describe the Hamiltonian dynamics of closed filaments. 
In the author's opinion, such a modification would make it possible to describe the dynamics of vortices in a turbulent flow.  We won't go into too much detail here, but we will focus on describing a single perturbed vortex ring with a finite radius R from a quantum perspective.

 Let the vector  ${\boldsymbol r}(\tau,\xi)$ satisfies the equation (\ref{LIE}).  
Evolution parameter $\tau$ was defined here as $\tau = {t\Gamma  }/{4\pi R^2}$, where $t$ is real time and symbol $\Gamma$ means the circulation. 
Let's note that we have redefined the evolution parameter for the curve (\ref{proj_r}) only. Any physical quantity (for example, circulation $\Gamma$) that is not associated directly  with curve (\ref{proj_r}) as a geometric object, remains unchanged.
Dimensionless parameter $\alpha$ depends on the filament core  radius $a<<R$ as well as  some details of regularization procedure that has been fulfilled  to deduce equation (\ref{LIE})
(see, for example, \cite{Saffm,AlKuOk}).
In the future, we will exclude this parameter from consideration by redefining of the parameter
$\tau$:  $\tau \to \alpha\tau$
Then, the function ${\boldsymbol j}(\tau,\eta)$ satisfies the equation for continuous Heisenberg spin chain:
\begin{equation}
        \label{CHSCeq}
        \partial_\tau {\boldsymbol{j}}(\tau ,\xi) ~=~ 
   {\boldsymbol{j}}(\tau ,\xi)\times\partial_\xi^{\,2}{\boldsymbol{j}}(\tau,\xi)\,.
        \end{equation}

		\section{Classical dynamic system}	
			
			\paragraph~
			
			Our subsequent purpose is to construct a quantum description of the dynamical system in question.   				The quantization of  non-linear integrable  models is complex problem (see, for example,  	\cite{TakFad_Q,Sklyanin}). 
			That is why 	we consider the small ($\varepsilon << 1$)  perturbation  ${\boldsymbol{j}}_p(\tau,\xi)$ of  the vortex ring  	${\boldsymbol{j}_0}(\xi)$ only: 
 		\begin{equation}
        \label{tang_v_per}          																
		\boldsymbol{j}(\tau,\xi)  ~=~ {\boldsymbol{j}_0}(\xi) + \varepsilon {\boldsymbol{j}}_p(\tau,\xi) \,,	\qquad \qquad		{\boldsymbol{j}}_p(\tau,\xi + 2\pi) = {\boldsymbol{j}}_p(\tau,\xi)\,,													
\end{equation}
where  unperturbed  vector $\boldsymbol{j}_0(\xi)$ is a tangent vector for the circular filament:
\[ \boldsymbol{j}_0(\xi) ~=~ \boldsymbol{e}_\phi   \,.  \]
The vectors $ ( {\boldsymbol{e}_\rho}\,,     \boldsymbol{e}_\phi\,, {\boldsymbol{e}_z}\,)$ denote the local cylindrical basis that is  associated naturally with this circular filament.  
	Please note that  this basis  differs from 
the basis $(\,\boldsymbol{n}_\rho\,, \boldsymbol{n}_\varphi\,,  \boldsymbol{n}_z\,)$  which was previously associated with domain $D$.

	Additionally, we  suppose that the excitations  (\ref{tang_v_per}) be transverse:
\begin{equation}
        \label{iden_1}
				{\boldsymbol j}_p(\tau ,\xi){\boldsymbol{j}_0}(\xi) ~\equiv~ 0\,
	\end{equation}
The small perturbation of the circular vortex filament  without the restriction (\ref{iden_1}) has been studied in the work \cite{Ricc}.
	Various aspects of the theory related to vortex ring oscillations have also been studied in the works \cite{Kop_Chern,Kik_Mam}.  The small perturbations of straight vortex filaments  were studied in the work \cite{Majda}.

	Condition (\ref{iden_1}) leads to decomposition
\begin{equation}
        \label{j-two}
{\boldsymbol j}_p(\tau ,\xi)   
  ~=~   j_\rho(\tau,\xi){\boldsymbol{e}}_\rho  +       j_z(\tau ,\xi)  {\boldsymbol{e}_z}\,,
\end{equation}

		For convenience, we  define the  complex-valued function
	\[  {\mathfrak J}(\tau ,\xi)  ~=~  j_\rho (\tau ,\xi) + {\rm i}  j_z(\tau ,\xi)\,.\]
	This function allows us define the complex amplitudes ${\mathfrak j}_{\,n}(\tau)$:
		\begin{equation}
        \label{sol_general}
				{\mathfrak J}(\tau ,\xi) ~=~ \sum_{n} {\mathfrak j}_{\,n}(\tau)\, e^{\,i\,\,n\xi }  \,.   
\end{equation}	
	
	Taking into account the terms of the order $\varepsilon$ only, we can deduce that 
	amplitude ${\mathfrak J}(\tau ,\xi)$
	satisfies the linear equation  \cite{Tal_EJM} 
		\begin{equation}
        \label{lin_eq_com}
\partial_\tau {\mathfrak J} =  - i  \partial_\xi^{\,2} {\mathfrak J}  
  - \frac{i}{2}\Bigl({\mathfrak J}   - \overline{\,{\mathfrak J}\,}\, \Bigr)   \,.
\end{equation}
Consequently,	
	\begin{equation}
	\label{sol_1}
        				{\mathfrak J}(\tau ,\xi) ~=~ \sum_{n} {\mathfrak j}_{\,n}\, 
								e^{\,i\,[\,n\,\xi  +  n \sqrt{n^2 - 1}\,\tau\,]} \,,  
\end{equation}	
	where the   amplitudes  ${\mathfrak j}_{\,n}$  satisfy the  condition 
		\[\overline{\,	{\mathfrak j}\,}_{\,-n}     ~=~    2 \left[n\sqrt{n^2 -1} - n^2 + {1}/{2} \right]  {\mathfrak j}_{\,n} \,.\]
	 		Thus, we  declare  the position vector  $\boldsymbol{q}$ (see Eq. (\ref{proj_r}))    and complex amplitudes ${\mathfrak j}\,_{\,-n}$, where $n =  0,1,2, \dots$, as independent  variables.    These variables describe the dynamics of  small-perturbed  circle-like curve ${\boldsymbol{r}}(\tau ,\xi)$  in accordance with Eq.(\ref{LIE}).
	Note that we still describe  the  evolution of curve (\ref{proj_r}) as a formal geometric object only.
	According to our assumption,  the curve ${\boldsymbol{r}}(\tau ,\xi)$,  is a vortex filament. However, variables $\boldsymbol{q}$ and ${\mathfrak j}\,_{\,-n}$ do not take into account the movement of the surrounding fluid  any way.
	That is why we include the circulation $\Gamma$ as an additional independent variable in our theory.  
Therefore, we describe the dynamics of the vortex filament in our approach by 
the following set  ${\cal A}$ of the independent variables: 
\[  {\cal A} ~=~ \{\, \Gamma\,, \boldsymbol{q}\,, \boldsymbol{j}(\xi)\, 
 \}\,.\]
	Although these variables take into account the movement of the surrounding fluid in a minimal way, they are inconvenient for subsequent quantization.
	Therefore, we replace the set ${\cal A}$ by the set ${\cal A}^{\prime}$, where
\begin{equation}
\label{new_set}
{\cal A} ~\subset~
{\cal A}^{\,\prime} ~=~ \bigl\{\, {\boldsymbol p}\,,  {\boldsymbol{q}}\,;  {\boldsymbol j}(\xi) \,\bigr\}\,.
 \end{equation}
Inclusion  ${\cal A} ~\subset~
{\cal A}^{\,\prime}$ is performed by means of a well-known hydrodynamic formula
\begin{equation}
        \label{p_and_m_st}
        {\boldsymbol p} ~=~ \frac{\varrho_0}{2 }\,\int\,\boldsymbol{r}\times\boldsymbol{w}(\boldsymbol{r})\,dV\,,
        \end{equation}
where constant $\varrho_0$ means a fluid density and vector-function    $\boldsymbol{w}(\boldsymbol{r})$  means  the vorticity of the vortex filament.   This function  is calculated as
\begin{equation}
        \label{vort_w}
     {\boldsymbol{w}}(\boldsymbol{r}) ~=~  \Gamma
                  \int\limits_{0}^{2\pi}\,\hat\delta(\boldsymbol{r} - \boldsymbol{r}(\xi))\partial_\xi{\boldsymbol{r}}(\xi)d\xi\,.
       \end{equation}
 	Compact form for the formulas (\ref{p_and_m_st}) and (\ref{vort_w}) is more convenient.
	Taking into account Eq.(\ref{proj_r}), we can write:
	
\begin{equation}
        \label{impuls_def}
    {\boldsymbol{p}}  ~=~    \varrho_0 {R}^2 \Gamma        {\boldsymbol f} \,, 
							\qquad 
						{\boldsymbol f} ~=~ \frac{1}{2}\int\limits_{0}^{2\pi} \int\limits_{0}^{2\pi} \left[\, {\xi - \eta}\,\right]\,{\boldsymbol j}(\eta)\times{\boldsymbol j}(\xi)d\xi  d\eta\,.	
      \end{equation} 
It is clear that the replacement $\Gamma \to {\boldsymbol p}$ leads to the certain constraints on the set ${\cal A}^{\,\prime}$ in general. This issue has been investigated in the article \cite{Tal_EJM} in detail. 
 	For our subsequent purposes, we assume ${\mathfrak j}\,_{\,0} = 0$, just as ${\mathfrak j}\,_{\,-1} = 0$  (see Eq. (\ref{sol_1})). 
	As regards the equality ${\mathfrak j}\,_{\,-1} = 0$, this 
	 assumption makes independent  momentum variations $\delta {\boldsymbol p}$ and small  variations ${\boldsymbol j}_p$. Thus, the equality 
	\begin{equation}
        \label{p_Gamma}
	{\boldsymbol{p}}  ~=~    \pi\varrho_0 {R}^2 \Gamma        {\boldsymbol e}_z 
	\end{equation}
	is fulfilled in the case ${\mathfrak j}\,_{\,-1} = 0$.

	
	We are considering a non-relativistic theory;  therefore, it is natural to declare the Galilei group
	${\mathcal G}_3$ as the space-time symmetry group of our model. Moreover, we will consider the central extension 	$\widetilde{\mathcal G}_3$	 of this group here\footnote{We use one parameter  central extension.}. 					
		Corresponding central charge  $m_0$ (this constant is interpreted as a ''mass'') completes the list of fundamental constants of the theory. The other constants are as follows:	the fluid's density $\varrho_0$, the speed of sound in this fluid $v_0$. 	
	In order to simplify some formulas,  we will use the auxiliary  constants
$R_0 = \sqrt[3\,]{m_0/\varrho_0}$,  $~t_0 = R_0/v_0$ and   ${\cal E}_0  = m_0 v_0^2$ along with constants $\varrho_0$, $v_0$ and $m_0$.					
						
			The presence of the central extended Galilei 	group 			 allows us to apply				
the group-theoretical approach to the definition   of the energy of thin  vortex filament.
	Indeed,           the Lee algebra of the group $\widetilde{\mathcal G}_3$ has three Cazimir functions:  
	
	 \[ {\hat C}_1 = m_0 {\hat I}\,,\quad 
  {\hat C}_2 = \left({\hat M}_i  - \sum_{k,j=x,y,z}\epsilon_{ijk}{\hat P}_j {\hat B}_k\right)^2 
  \quad {\hat C}_3 = \hat H -  \frac{1}{2m_0}\sum_{i=x,y,z}{\hat P}_i^{\,2}\,,\]                         
       where        ${\hat I}$ is the unit operator,     ${\hat M}_i$,   $\hat H$,  ${\hat P}_i$         and  ${\hat B}_i$  ($i = x,y,z$)
        are the respective generators of rotations, time and space translations and Galilean boosts. 		 As it is well known, the function  ${\hat C}_3 $  can be interpreted as  an  ''internal energy of the particle''. 
	In our model, the internal degrees of freedom are described by the function ${\mathfrak J}(\tau ,\xi)$.  This function is  invariant under Galilei transformation; that is why ${\hat C}_3 = {\hat C}_3({\mathfrak J}) $ in general. Let us postulate a formula for the Cazimir functions ${\hat C}_3 $:
	
	 \[{\hat C}_3  			 ~=~
			{\mathcal E}_0 \sum_{n>1} |\,{\mathfrak j}_{\,-n}|^2 n\sqrt{n^2 -1}			\,.\]

	Therefore, the following expression for  the energy ${\mathcal E}$  is natural for our model:
	
	\begin{equation}
  \label{energy_1}
 {\mathcal E}({\boldsymbol p}\,;{\mathfrak j}\,) = \frac{{\boldsymbol{p}}^{\,2}}{2m_0}   +  
 {\mathcal E}_0 \sum_{n>1} |\,{\mathfrak j}_{\,-n}|^2 n\sqrt{n^2 -1}	\,.
\end{equation}
	
	This formula will be justified after determining the Hamiltonian structure of the theory.  The application of the group-theoretic approach allows us to avoid a detailed analysis of the structure  of the vortex filament    near thin core and the associated problems. Indeed, the size of both vortex loops and their cores may be in the range of angstroms.

	Let us define the Hamiltonian structure.
	
	\begin{itemize}
  \item The set ${\cal A}^{\,\prime}$ is interpreted as a phase space ${\mathcal H}$ quite naturally:   ${\mathcal H} =  {\mathcal H}_3  \times  {\mathcal H}_j   $. The space $ {\mathcal H}_3$ is  parametrized by the variables    ${\boldsymbol{q}}$ and  ${\boldsymbol{p}}$; this space is interpreted  as a  phase space of  $3D$  free structureless    particle.   The phase  space    $ {\mathcal H}_j$ is parametrized by the complex variables
	 	${\mathfrak j}_{\,-n}$,  $\overline{\,\mathfrak j}_{\,-n}$  ($n  =  1, 2, \dots$).		
		 \item Poisson structure:
  \begin{eqnarray}
	\label{pq_brackets}
	  \{p_i\,,q_j\} & = & \delta_{ij}\,,\qquad i,j = x,y,z\,,  \\[2mm]
  \label{ja_jb}
  \{ {\mathfrak j}_{\,m}, \overline{\,\mathfrak j\,}_{\,n}\} & = & (i/{\mathcal E}_0 t_0)\, \delta_{mn}\,, \qquad m,n = -1,-2,\dots
  \end{eqnarray}
  
	All other brackets  vanish. 		
	  \item Hamiltonian 
	\begin{equation}
	\label{H_ful}
	H({\boldsymbol p}\,;{\mathfrak j}\,)     ~=~  {\mathcal E}({\boldsymbol p}\,;{\mathfrak j}\,)\,,
	\end{equation}
	where  the function  ${\mathcal E}({\boldsymbol p}\,;{\mathfrak j}) $ was defined by the formula (\ref{energy_1}). 
  \end{itemize}

	Hamiltonian (\ref{H_ful}) defines the evolution of the filament ${\boldsymbol r}(\xi)$ in
	 ''conditional'' time $t^\# = t_0\tau$.
	It is not difficult to make sure that the Hamiltonian (\ref{H_ful}) and the Poisson brackets (\ref{pq_brackets}),  (\ref{ja_jb}) generate dynamics
	\begin{eqnarray}
	\label{q_tau}
	 {\boldsymbol q}(0) &\longrightarrow& {\boldsymbol q}(t^\#) ~=~ {\boldsymbol q}(0) ~+~ \frac{\boldsymbol p}{m_0}\,t^\#\,,\\[3mm]
	  {\mathfrak j}_{-n}(0)  &\longrightarrow& {\mathfrak j}_{-n}(t^\#) ~=~  {\mathfrak j}_{-n}(0)\,  e^{\,-i\,  n \sqrt{n^2 - 1}\,(t^\#/t_0)\,} \nonumber
	\end{eqnarray}
 Among other things,  these formulas demonstrate the fulfilled extension 
\[  \bigl\{\,  {\boldsymbol{q}}\,;  {\boldsymbol j}(\xi) \,\bigr\} ~\longrightarrow~
\bigl\{\, {\boldsymbol p}\,,  {\boldsymbol{q}}\,;  {\boldsymbol j}(\xi) \,\bigr\}\,\]
of the original dynamic system (\ref{LIE}).
Indeed, in accordance with Eq.(\ref{LIE})
\[ \frac{d {\boldsymbol q}}{d t^\#} ~=~  \frac{R}{t_0}\,{\boldsymbol e}_z\,.\]
	As opposed to Eq.(\ref{q_tau}),   this formula corresponds to the following dependence ${\boldsymbol q} = {\boldsymbol q}(t^\#)$:
		\[ {\boldsymbol q}(0) ~\longrightarrow~ {\boldsymbol q}(t^\#) ~=~ {\boldsymbol q}(0) ~+~ 
		{R}\,(t^\#/t_0)\,{\boldsymbol e}_z\, .\]

	\section{Quantization and energy spectrum}

	\paragraph~
	
	Let's quantize  the constructed dynamic system.
											Firstly, we  define a Hilbert space $\boldsymbol{H}$   of the quantum states.   The structure of the phase space ${\mathcal H}$ makes the following construction natural:
																			
					\begin{equation}
	\label{space_quant}
	\boldsymbol{H}  =  \boldsymbol{H}_3 \otimes   \boldsymbol{H}_j\,,
	\end{equation}
			where the symbol   $\boldsymbol{H}_3$  denotes the Hilbert space  of a free structureless particle in the space  ${\sf R}_3$ (the space 			
			$L^2({\sf R}_3)$ in our case) and  symbol $\boldsymbol{H}_j $ denotes the Fock space for the infinite number of the harmonic oscillators.
		This space is formed by the basic vectors	
		\[  \bigl(\hat{a}^+(n_1)\bigr)^{s_1} \dots \bigl(\hat{a}^+(n_N)\bigr)^{s_N}|0\rangle\,, \qquad n_i \not= n_j \quad for \quad i \not= j\,,\]
		where  oscillators are numbered with numbers $n_1,\dots,n_N$, 
				 $N = 0,1,2,\dots$ and  symbols $s_j$ denote  corresponding occupation numbers. Creation and annihilation 	
			 operators in the space 	$\boldsymbol{H}_j $   have standard commutation relations
				\[ [\,\hat{a}(m), \hat{a}^+(n)] = \delta_{mn}\hat{I}_j\,, \qquad \hat{a}(m)|\,0\rangle = 0  \,, \qquad  m,n = 1,2,\dots \,, \qquad    |\,0\rangle \in    \boldsymbol{H}_j           \,,\]   
				where the operator   $\hat{I}_j$  is unit operator in the space    $\boldsymbol{H}_j $.
	
		Our postulate of quantization is following:
					\[ \boldsymbol{q} ~\to~  \boldsymbol{q}\otimes \,\hat{I}_j  \,,\qquad  \boldsymbol{p}  ~\to~ - i\hbar \nabla \otimes \,\hat{I}_j \,,\qquad
					{\mathfrak j}_{\,-n} ~\to~ \sqrt{\frac{\hbar}{t_0{\mathcal E}_0}}\, \Bigl(\hat{I}_3 \otimes\,\hat{a}(n)\Bigr)\,, \]
					where  $ n = 1,2,\dots\, $.  The unit operator $\hat{I}_3$ acts in the space $L^2({\sf R}_3)$.  In order to downsize  the    formulas,
  we will not write the constructions $(\dots  \otimes \,\hat{I}_j)$  and  $ (\hat{I}_3 \otimes\,\dots)$ explicitly, hoping that this simplification will not lead to misunderstandings.
	As usual in a field theory, we postulate the normal ordering for the operators  $\hat{a}(m)$ and $\hat{a}^+(n)$ when we quantize any products of the variables ${\mathfrak j}_{\,-n}$  and  $\overline{\,{\mathfrak j}\,}_{\,-m}$. For example,
	\[  |{\mathfrak j}_{\,-n}|^2 ~\to~  \frac{\hbar}{t_0{\mathcal E}_0} \hat{a}^+(n)\hat{a}(n)\,.\]

	In this study  we  consider our system  in the domain (\ref{domain}) only.
	 			Before calculating  the energy spectrum, we quantize circulation $\Gamma$.
		Let's pay attention that our approach allows us to find the spectrum of the value $\Gamma$ using conventional methods of quantum theory. 
	Arguments\footnote{In addition to the arguments given at the beginning of the article.} in favor of generalizing the standard expression for quantum circulation were discussed in more detail in the author's article \cite{Tal_PhRF23}.

	Squaring the Eq. (\ref{p_Gamma}), we have $SO(3)$ - invariant expression	
		\[  \boldsymbol{p}^2 ~-~  \pi^2 \varrho_0^2 R^4 \Gamma^2 ~=~ 0\,.\]
		Therefore, quantized circulation values are the eigenvalues of the spectral problem in the space $L^2({\sf R}_3)$:
		\begin{equation}
	\label{eq_sp_Gamma}
	\left[ - \hbar^2 \Delta  ~-~  \pi^2 \varrho_0^2 R^4 \Gamma^2 \,  \right]\, \Psi_\Gamma(\boldsymbol{r})  ~=~ 0\,, \qquad  \Psi_\Gamma(\boldsymbol{r}) \in L^2({\sf R}_3)\,.
	\end{equation}
		Solving this simple spectral problem, we find the spectrum of the value $\Gamma$:
		\begin{equation}
	\label{spectr_Gamma}
		\Gamma_{n,m,k} ~=~  \frac{\hbar }{\pi \varrho_0 R^2} \sqrt{~\biggl(\frac{\pi n}{L}\biggr)^2  + \biggl(\frac{\zeta_{\,k}^{(m)}}{R_1}\biggr)^2~}\,, \qquad n,m = 1,2,\dots, \quad k = 0,1,2,\dots\,.
	\end{equation}	
		The number $\zeta_k^{(m)}$ means  $m$-th zero of the Bessel function $J_k(x)$.
				
				Let us investigate some limiting cases for the formula (\ref{spectr_Gamma}).
		\begin{enumerate}
		\item The value $L$ (''length of the pipe'') is finite and be  comparable with the pipe radius $R_1$, quantum numbers $n$ and $k$ are small. 
		Because the asymptotic behavior
		\begin{equation}
		\label{zeta_asympt}
		\zeta_k^{(m)} ~\longrightarrow~  -\frac{\pi}{4} + \frac{\pi k}{2}  +\pi m \,, \qquad m \to \infty
		\end{equation}
		takes place, the formula (\ref{spectr_Gamma}) has the following asymptotic behavior for large values of the number $m$:
		\[ \Gamma_{n,m,k} ~\longrightarrow~  \Gamma_{m}  ~=~ \frac{\hbar\, m}{ \varrho_0 R^2 R_1} \,.\]
		\item The value $L$ is very large, quantum numbers $n$, $m$ and $k$ take arbitrary but finite values.
			In this case circulation $\Gamma$ takes the values within   the bounded interval
			$(\,\Gamma_{min}, \Gamma_{max}\,)$. Obviously,   
			\[\Gamma_{min} = 			\frac{\hbar \zeta_{\,0}^{(0)}}{\pi \varrho_0 R^2 R_1}\,.\]
			The gap between adjacent values is as follows:
			\begin{equation}
			\label{gap}
			  \delta_n\Gamma_{n,m,k} ~\equiv~ \Gamma_{n+1,m,k} - \Gamma_{n,m,k} ~=~ 
				\frac{\pi \hbar R_1}{2\varrho_0 R^2}\,\frac{2n+1}{\zeta_{\,k}^{(m)}L^2} ~+~ 
				{\cal O}\left( \frac{1}{L^4}\right)\,.
			\end{equation}
					
						To define the value $\Gamma_{max}$, we postulate that the vortex momentum $\boldsymbol{p}$ is bounded in our theory:
			\[   |\,\boldsymbol{p}\,| ~<~   m_0 v_0\,.\]
			This restriction  can be justified by the fact that we consider only subsonic fluid movements.
			Taking into account Eq. (\ref{p_Gamma}), we deduce
			\begin{equation}
	\label{Gamma_max}
			\Gamma_{max} ~=~  \frac{m_0 v_0}{\pi\varrho_0 R^2} ~=~ \hbar\frac{{\sf k}_{max}}{\pi\varrho_0 R^2} \,,
			\end{equation}
			where the number ${\sf k}_{max} = m_0 v_0/\hbar$ is maximal De Broglie wave number.
			Thus, we have following formula for the case $L\to\infty$:
			\begin{equation}
	\label{Gamma_con}
			\Gamma ~=~ \frac{\gamma \,\hbar}{\pi \varrho_0 R^2 R_1}\,,
						\end{equation}
			where the real number $\gamma \in (\,\zeta_{\,0}^{(0)}\,, {\sf k}_{max}R_1)$.
			Formula (\ref{Gamma_con}) allows us to interpret the number $\gamma$  as
			a  ''dimensionless quantum circulation''.   			We also assume that 
			$\zeta_{\,0}^{(0)}  << {\sf k}_{max}R_1$   in this model.
		\end{enumerate}
		
		Our next step is to find the energy spectrum of the dynamical system under consideration.
		Here we need to remember that energy is a physical value that is associated with time translations.
		Therefore,  we must return to  of evolution of the vortex loop in real time $t$ instead of ''conditional'' time $t^\#$:
		\begin{equation}
	\label{t_real}
		 t^\# ~\to~  t ~=~ \frac{4\pi R^2}{\alpha\, t_0 \Gamma}\,t^\#\,.
		\end{equation}
		First, we consider the $t^\#$ - evolution.
		Quantized Hamiltonian (\ref{H_ful}) has following form:
		\begin{equation}
	\label{H_quant}
		\widehat{H}  ~=~ -  \frac{\hbar^2}{2m_0}\Delta ~+~ \frac{\hbar}{t_0}
		\sum_{n>1}  n\sqrt{n^2 -1}\,\hat{a}^+(n)\hat{a}(n)\,.
		\end{equation}

		Spectral problem
		\[\widehat{H} |\Psi\rangle ~=~ E^\#|\Psi\rangle\]
		has following solutions:
	\[ \boldsymbol{H} \ni   |\Psi\rangle ~\equiv~  |\Psi(\Gamma;\ell, s_\ell)\rangle ~=~ 
	|\Psi_\Gamma\rangle|\ell, 
		s_\ell\rangle\,, \qquad\quad
		|\ell, s_\ell\rangle = \bigl( \hat{a}^+(\ell) \bigr)^{s_\ell}|0\rangle \in \boldsymbol{H}_j\,,\] 
		where  the vector $|\Psi_\Gamma\rangle \in  L^2({\sf R}_3)$  is the eigenvector of the spectral problem
		(\ref{eq_sp_Gamma}) that corresponds to certain  eigenvalue $\Gamma$.
				Taking into account formula (\ref{eq_sp_Gamma}),
		the eigenvalues $E^\#_{nmk}(\ell, s_\ell)$ are written as follows: 
		\begin{equation}
	\label{E_eigen1}
	E^\#_{nmk}(\ell, s_\ell) ~=~  \frac{(\pi\varrho_0 R^2)^2}{2m_0}\Gamma_{n,m,k}^{\,2} ~+~
	\frac{\hbar}{t_0}\,  \ell\sqrt{\ell^2 -1}\,s_\ell\,,
	\end{equation}
	where numbers $\Gamma_{n,m,k}$ have been defined in accordance with formula (\ref{spectr_Gamma}).
		Next, we will consider the case $L \to \infty$ only. 
		  		Since this case has been investigated above (see Eq.(\ref{Gamma_con}), 
				we can write down the formula for the energy
		$E^\#(\gamma;\ell, s_\ell) = \lim_{L\to \infty}E^\#_{nmk}(\ell, s_\ell)$:
		\begin{equation}
	\label{E_eigen2}
	E^\#(\gamma;\ell, s_\ell) ~=~  \frac{\gamma^2 \hbar^2}{2m_0R_1^{\,2}} ~+~
	\frac{\hbar}{t_0}\,  \ell\sqrt{\ell^2 -1}\,s_\ell\,,\qquad \gamma \in (\zeta_{\,0}^{(0)}\,, {\sf k}_{max}R_1)\,.
	\end{equation}
		Let us consider the $t^\#$ - evolution of any vector $|\Psi\rangle \in {\boldsymbol H}$:
		\begin{equation}
	\label{t_ev1}
		\exp\left(\frac{i\widehat{H} t^\#}{\hbar} \right)|\Psi\rangle ~=~  
		\sum_{\Gamma;\ell, s_\ell}C_{\gamma;\ell, s_\ell}\exp\left(\frac{iE^\#(\gamma;\ell, s_\ell) t^\#}{\hbar} \right)|\Psi(\Gamma;\ell, s_\ell)\rangle\,,
		\end{equation}
	where  $ C_{\gamma;\ell, s_\ell} = \langle\Psi(\Gamma;\ell, s_\ell)|\Psi\rangle$.	
				Next, we  restore the real time $t$ in this formula using Eq.(\ref{t_real}).


		\begin{figure}[t]
		\label{spectrum_1}
{ {\hspace{20mm}~\includegraphics[width=3.5in]{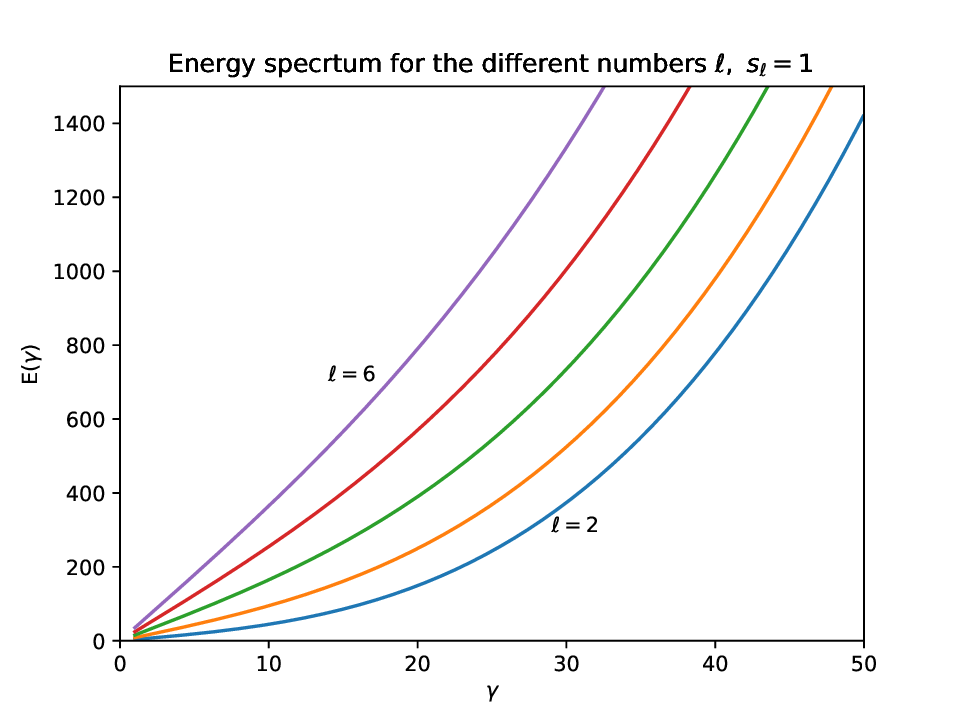}}}
{\vspace{70mm}\caption{The energy spectrum for the case $L\to\infty$.}}
\end{figure}

				We assume that following equality takes place for the real energy $E(\gamma;\ell, s_\ell)$: 
				\[  E^\#(\gamma;\ell, s_\ell) t^\#  ~=~  E(\gamma;\ell, s_\ell) t\,.\]
				Therefore, 				the ''real - time'' evolution of any vector $|\Psi\rangle \in {\boldsymbol H}$   is written as
		\[ |\Psi\rangle ~\longrightarrow~ |\Psi(t)\rangle ~=~
		\sum_{\Gamma;\ell, s_\ell}C_{\gamma;\ell, s_\ell}\exp\left(\frac{iE(\gamma;\ell, s_\ell)\,t}{\hbar} \right)|\Psi(\Gamma;\ell, s_\ell)\rangle\,, \]	
					where 
		
		\begin{equation}
	\label{E_final}
		E(\gamma;\ell, s_\ell) ~=~ \frac{\alpha \,\hbar^2}{4\pi^2 \varrho_0 R^4 R_1}\left( \frac{R_0}{2R_1}\,\frac{\gamma^3}{{\sf k}_{max}R_1} ~+~ \gamma\, \ell\sqrt{\ell^2 -1}\,s_\ell  \right)\,.
		\end{equation}
		
		Finally, the numbers $E(\gamma;\ell, s_\ell)$ give the  energy spectrum of our quantum system.
		The dependence $E = E(\gamma;\ell, s_\ell)$ is shown in Fig.1 for some conventional scale units.
		
		Of course, the limit $L = \infty$  is unattainable really. 
		Therefore, the ''dimensionless  circulation'' $\gamma$    takes on a value in  discrete set $G_\epsilon =(\gamma_1,   \gamma_2, \dots,   \gamma_n, \dots )$  only. This set $G_\epsilon $ is the $\epsilon$-net for the interval $(\zeta_{\,0}^{(0)}\,, {\sf k}_{max}R_1)$. In accordance with formulas (\ref{gap}) and (\ref{Gamma_con})  parameter $\epsilon$ that characterizes the set $G_\epsilon$ is defined as follows:
	\[ \epsilon ~=~  \epsilon(m,n,k) ~=~   \frac{\pi^2}{2}\, \frac{2n+1}{\zeta_{\,k}^{(m)}}\left(\frac{R_1}{L}\right)^2\,.\]
		Thus, the curves that were shown  in Fig.1 are not continuous. Really, these curves  consist   of many nearby points. 
				In accordance with asymptotic (\ref{zeta_asympt}), the numbers  $\epsilon(m,n,k)$  demonstrate the structure of a natural series  for large  quantum number $m$ (numbers $n$ and $k$ are fixed).
		Therefore, the set of the numbers $\Gamma$ (as well as the set of the numbers $E$)  
				has the fractal structure: this structure is transferred from the fractal structure of the natural series		 $1/m$
in a neighborhood of the point $(1/m) \to 0$.

		To sum up this section, we note that all the results were obtained under the assumption that the quantities $R$, $R_1$, $L$ and $m_0$ are finite.  
	Some of the definitions made above are valid in this case only.
				The proposed theory, we believe, is  applicable to the quantum description of meso- and  microscopic vortex loops.
		The limit  $R \to \infty$ refers to the transition from a microscopic to a macroscopic object.
		In general, these objects are described by classical theory.
		To describe the considered dynamical system in this limit, we need to use different methods and make additional assumptions.

		\section{Concluding remarks}

		\paragraph~
		
	As it seems, 	 proposed quantum system is ''inverse realization''  of Lord Kelvin's old idea that any particles can be considered as some kind of vortex-like structures  \cite{Thom}. In our approach, we describe the closed vortex filament  as some structured particle. 
	In this regard, graphs of the function $E = E(\gamma)$ for the different numbers $\ell$ and $s_\ell$  can be considered as some kind of analogue of Regge trajectories in the theory of elementary particles.  Why can the dependence $E = E(\gamma)$ be called a Regge trajectory?
	Indeed, the conventional formula for the fluid angular  momenta  as follows \cite{Saffm}
	\[\boldsymbol{s} =\frac{1}{3}\, \int\,\boldsymbol{r}\times
         \bigl(\boldsymbol{r}\times\boldsymbol{\omega}(\boldsymbol{r})\bigr)dV\,,\]
		where the vorticity $\boldsymbol{\omega}$ was defined by the expression (\ref{vort_w}).
	Therefore, 
	\[ |\boldsymbol{s}| ~\propto~ \Gamma ~\propto~ \gamma\,.\] 
			In a sense, the fluid in which the vortex  in question moves   plays the role of ''aether''.
	Note that such  ''aether'' is quite real here, and not hypothetical.

	In this study, we applied a group-theoretical approach to the definition of energy.  Unfortunately, the author did not find any experimental studies, the results of which could be compared with the formula (\ref{E_final}).
	Hopefully, such studies will  appear in the future.

	\end{document}